# Data-driven Innovation: Understanding the Direction for Future Research

## Research-in-progress

### Sasari Samarasinghe
School of Business
University of Southern Queensland
Springfield, Queensland, Australia
Email: udanjala@gmail.com

### Sachithra Lokuge
School of Business
University of Southern Queensland
Springfield, Queensland, Australia
Email: ksplokuge@gmail.com

## Abstract

In the contemporary age of information, organisations have realised the importance of "data" to innovate and thereby attain a competitive advantage. As a result, firms are more focused on understanding the potential to achieve data-driven innovation (DDI).  Researchers too have focused on examining this novel phenomenon in a broader scope. In this study, we conducted a systematic and comprehensive review of the literature to understand the DDI phenomenon. The findings of this study benefit scholars in determining the gaps in the current body of knowledge as well as for practitioners to improve their data strategy to enhance and develop innovation capabilities.

## 1   Introduction

Data is considered as the engine of the global digital economy. To harness the true power of data and analytics, organisations should formulate a secure data strategy that broaden the access to data and promote a data culture that enables data monetisation and data-driven innovation (Legner et al. 2020; Shi et al. 2022; Symons 2022). With the growing interest in investing in data initiatives, organisational data is recognised as a strategic resource and a catalyst for developing innovation capabilities (Lokuge and Sedera 2020; Sedera and Lokuge 2019a). At present, data-driven innovation (DDI) is considered a key pillar of the business growth (Chatterjee et al. 2022). In this paper, data-driven innovation is defined as *"a strategic initiative of organisations to use data and analytics to develop data-driven insights that help for new product development, process improvements, discover new markets and business models"* (OECD 2015; Su et al. 2022). Technological advancements such as the internet-of-things (IoT), big data capabilities, cloud computing and other technologies have reduced the cost of data collection, storage (Sedera et al. 2022a), and processing (Sedera and Lokuge 2018), expanding the opportunity for organisations to harness the value of data and promote innovation (Chatterjee et al. 2022; Del Giudice et al. 2020). Organisations can extend their existing innovation skills, processes, and knowledge to improve innovation competencies or renew their exploration competencies by implementing data-driven strategies (Adikari et al. 2021; Sedera and Lokuge 2017; Visvizi et al. 2021).

While DDI is popular among many firms and firms are investing extensively in data initiatives to create an environment in which data drives decisions, academic research on this area is still growing and scattered. Several studies have shown that technological advancements and active support from the leadership promote the development of data-driven insights and innovation capabilities (Chatterjee et al. 2022; Ghasemaghaei and Calic 2019; Lokuge et al. 2019). As such, the objective of this paper is to review the existing literature on DDI and understand how organisations have used data and analytics tools to facilitate innovation processes. Thereby, we aim to identify future research directions to guide future academic work towards a more coherent and robust understanding of how organisations can use data-driven insights in their innovation process. To achieve this goal, we used two themes to fragment the findings from prior research: (1) antecedents of data-driven innovation and (2) outcomes of data-driven innovation. In doing so, we summarise existing studies on important enablers and outcomes of data-driven innovation and identify promising future research opportunities.

The remainder of the paper is structured as follows: The next section explains the methodology and analysis process. Then, the results of the systematic literature review are provided next. The conclusion section provides academic contributions, practical contributions, limitations, and future directions.

## 2   Research Methodology

To analyse and understand the current state of data-driven innovation research, we conducted a systematic literature review following the guidelines of Levy and Ellis (2006). The study considered forty-one (41) Scimago Q1 and Q2 ranked journals in 10 online databases (i.e., AIS virtual library, Emerald insight, Springer, Springer Open online library, ScienceDirect, ACS Journals, SAGE, Informs online library, Taylor & Francis online library, and Wiley online library). We also referred to the Australian Business Deans Council (ABDC) journal list and filtered the journals related to the information systems (IS) and management disciplines. Keywords used for the search string were "data-driven innovation" OR "data-enabled innovation" OR "big data innovation" OR "data-infused innovation" OR "data innovation" OR "data-based innovation."

The search result included one hundred and sixty-one (161) journal articles. Since most of the research studies through the search were published in the last 4-5 years, for the analysis, we considered ten years period from 2013-2022. Also, only the journal articles written in the English language were considered for the study. After reviewing each journal article, we excluded the articles unrelated to the research objective, or that did not fit with the research topic. Next, similar to Sedera et al. (2017) duplicates were removed, and the final paper sample filtered for the analysis included twenty-three (23) journal papers. Most of the journal articles used for the analysis were published in Q1 journals (i.e., Annals of Operations Research, Journal of Product Innovation Management, Journal of Strategic Information Systems, Journal of Business Research, European Journal of Innovation Management, Information Systems Frontiers, and Journal of Big Data). Each paper was read and documented important information on how data-driven innovation has been defined, its limitations and future research scope. Table 1 below outlines the selected papers for the literature review.

| Year | References | No. of Papers |
|---|---|---|
| 2022 | (Bhatti et al. 2022), (Shi et al. 2022), (Su et al. 2022) | 03 |
| 2021 | (Babu et al. 2021), (Visvizi et al. 2021), (Saura et al. 2021), (Bresciani et al. 2021), (Akter et al. 2021), (Kozak et al. 2021), (Chatterjee et al. 2021), (Belhadi et al. 2021), (Chaudhuri et al. 2021) | 09 |
| 2020 | (Rizk and Elragal 2020) | 01 |
| 2019 | (Ghasemaghaei and Calic 2019), (Trabucchi and Buganza 2019), (Bharati and Chaudhury 2019) | 03 |
| 2017 | (Chandy et al. 2017), (Troilo et al. 2017), (Janssen et al. 2017), (Dwivedi et al. 2017), (Sorescu 2017) | 05 |
| 2015 | (Bosch-Sijtsema and Bosch 2015) | 01 |
| 2013 | (Provost and Fawcett 2013) | 01 |

*Table 1. Papers Selected for the Literature Review*

## 3  Analysis and Discussion

Based on the literature review, the authors identified antecedents and outcomes of the data-driven innovation process. Our objective during the analysis was to understand the data-driven innovation process and future research directions. From the sample, it was evident that many studies were conducted in the UK and Europe, focusing on product innovations in the manufacturing sector, supply chain innovation, process innovation, and public sector innovation. The studies have used the theoretical foundation of the resource-based view (RBV), dynamic capability theory, critical-mass theory, information overload theory, institutional theory, and TOE framework according to their research design. The selected papers have used a fair distribution of qualitative case studies and quantitative survey methods for data collection and analysis. In contrast, some journal articles have used systematic literature review and bibliometric analysis techniques in their studies (around 15% of the papers). Also, it was noted that almost all studies were conducted in a single country, indicating the lack of cross-country comparisons.

### 3.1  Antecedents of Data-driven Innovation Process

According to the analysis results, various factors influence the data-driven innovation process at the organisational and individual levels. At the organisation-level, previous studies have identified factors such as robust analytical capabilities, moving to a data-driven culture, technological capabilities, stakeholder involvement, incentives for stimulating innovation, strong data governance, data privacy and access regulations as influencing factors for the successful implementation of DDI process (Babu et al. 2021; Bhatti et al. 2022; Chandy et al. 2017; Janssen et al. 2017; Shi et al. 2022). An organisation's decision to move with data-driven innovation is a costly investment (Lokuge et al. 2020). Therefore, making necessary human resources available, top management support, budget and other innovation means were considered essential (Janssen et al. 2017; Lokuge and Sedera 2014a; Palekar et al. 2013). At the same time, technological readiness and analytical capabilities are imperative to collect and share data, combine data, and analyse data, which are specific to the data-driven innovation (Babu et al. 2021). Prior researchers have discussed the importance of changing traditional siloed culture to a data culture as an influencing factor for data-driven innovation. By moving so, culture enables organisations to quantify risks in monetary terms rather than being risk-averse and encourages collaboration with others (Alarifi et al. 2015; Janssen et al. 2017; Lokuge and Sedera 2016). Data-driven innovation also requires access to data and the ability to reuse and share data (Shi et al. 2022). One of the organisations' main challenges in data utilisation is the privacy concern involved in the data collection and aggregation (Ghasemaghaei and Calic 2019). Therefore, previous studies have highlighted the importance of solid data governance as another influencing factor for the data-driven innovation (Babu et al. 2021; Janssen et al. 2017).

At the individual level, previous studies have identified factors such as the capabilities and knowledge of employees to innovate using the data as enablers for the innovation process (Ahmad et al. 2013; Janssen et al. 2017). Organisations can follow a top-down skills management plan (i.e., define job roles and the skills required to perform those roles), aligning with the overall organisation strategy (Shi et al.

2022) to develop and improve the knowledge and analytical skills of employees. However, most research studies have focused on organisation-level capabilities, and only limited studies have discussed individual-level capabilities. Therefore, this is an area that requires further investigation. Based on the analysis, several future research suggestions were identified related to exploring antecedents of data-driven innovation. Further research can be carried out to identify external factors that can influence the innovation process and the effect of individual factors such as motivation, demographic variables, and job characteristics in the prediction of data-driven innovation that promote data-driven decision-making, creative self-efficacy and creative performance (Babu et al. 2021). Analysis of the social processes connected to innovation in a data-driven environment is another research area that academic researchers can examine how social processes like conflict and power shifts can influence the innovation process (Troilo et al. 2017). Future studies can also assess the relationship between stakeholders and the analytic capabilities of the organisations that influence the innovation capabilities, as well as how big data can influence an organisation's innovation competency for more specific types of firms such as multinational corporations, small and medium-sized enterprises (Bhatti et al. 2022; Ghasemaghaei and Calic 2019).

### 3.2  Outcomes of Data-driven Innovation Process

Based on the results, the authors identified two categories of outcomes: economic and social outcomes. Organisations should re-focus their strategic objectives to consider data as the key levers of efficient decision-making and implement smart technologies for efficient and cost-effective data management practices that drive the conversion of data into data-driven insights (Visvizi et al. 2021). The ability to integrate, analyse and interpret data to produce data-driven insights should align with the strategic objectives. Support from top management and supervision by managers to enable the transformation of data into knowledge is another essential practice that can help to achieve benefits from data-driven innovation practices (Ghasemaghaei and Calic 2019; Lokuge and Sedera 2014b). Based on the analysis, we identified economic benefits that include introducing new products and services that simplify and boost business processes, increase efficiency, and effectiveness of operations, increase customer satisfaction, improve organisational performance, and ultimately achieve a competitive advantage over rival firms in the market. Once an organisation establishes and implements the correct data strategy to enhance innovation activities, it will improve product development opportunities and organisational performance (Chaudhuri et al. 2021; Sedera et al. 2022b). Firms often use key performance indicators (KPIs) such as research and development (R&D) expenditure and growth indicators (i.e., sales growth, number of products, number of new customers) as innovation capability measures to assess their success in innovation activities (Borocki et al. 2013; Chaudhuri et al. 2021; Sedera 2006; Sedera and Lokuge 2019b), and this could be a future research area to determine how firms can measure data-driven innovation capabilities. Improving the organisation's creative ability with data and analytics would help develop new products that meet the market demand. However, the external environment could be a moderate factor; therefore, organisations should understand the market's needs to improve their capability in the product innovation (Chatterjee et al. 2021). Adapting to business analytics and accuracy in data analysis will also help organisations respond to external market changes. We believe future research is essential as fewer studies have focused on this direction.

Based on the analysis, we identified several social benefits, such as employee motivation (including ground-level staff), empowerment, work efficiency, and accountability (Janssen et al. 2017). When employees (especially the ground-level non-technical staff) are given access to data, provided with the right analytical tool(s) and provided an environment to improve data analytical skills and data awareness (i.e., a democratized data environment), organisations can improve employees' innovative thinking and creativity (Ghasemaghaei and Calic 2019; Lefebvre et al. 2021). This strategic move empowers employees by giving true ownership of their work while simultaneously building trust and credibility on data to develop insights for solving operational challenges (Janssen et al. 2017; Lokuge and Sedera 2018). Therefore, adequate training on data awareness and improving analytical skills are equally essential to ensure that innovative moves or actions taken by employees are accurate, efficient, and effective.

## 4   Conclusion

Although there is a growing interest in data-driven innovation research, the overall body of research is still under-developed. We identified 23 journal articles in IS and management disciplines based on the analysis. The results were categorised into two themes to identify antecedents and outcomes of the data-driven innovation process and future research avenues. Under each category, the main findings and future research suggestions are identified through the review process. Based on the analysis, we found that product and process innovation types are widely studied as less attention has been given to the business model and market innovation. This could be a good area for future studies as it links with the competitive advantage. Studying how organisations can develop data-driven insights through data management practices such as "data democratization" is another potential area for further study. Data democratization is a growing research area that helps organisations for empowering employees to use data to drive decision-making and foster innovation. However, how it helps in the innovation process is still understudied. Also, studying the dark side of the data-driven innovation process will be a value-adding area allowing organisations to learn from failures and improve their innovation process. For example, areas like how inappropriate data analytical tools can add up to a competitive disadvantage to digitalize the organisation and innovation activities could also be a future research area that would add value to the academia (Chaudhuri et al. 2021; Lokuge and Duan 2021; Lokuge et al. 2016). Through the analysis, we found that the outcomes of data-driven innovation have gained little attention in the literature. Therefore, future research has the potential to examine these economic and social outcomes as measures of the innovation process. An in-depth study on how new products or services will be incorporated into the business models and whether business model innovation is a more potent contributor than product innovation to achieve competitive advantage would also be an attractive study area for future work (Sedera et al. 2016a; Sorescu 2017). Data can strengthen the competitive advantage by offering real-time data-driven insights (i.e., through data management practices like data democratization). These insights can help organisations update their business models according to changing market conditions. Other areas for future research include assessing the relationship between data-driven decision orientation and innovative outcomes, considering different actors' engagement, users' behaviour, and value co-creation in the innovation process (Lokuge et al. 2018; Sedera et al. 2016b; Visvizi et al. 2021).

We understand that the sample size was small. However, an in-depth literature analysis was done to identify the current state of data-driven innovation research and future research directions. Also, the findings of this study provided an understanding of existing knowledge and IS research areas on data-driven innovation. We look forward to extending our research to study how data-driven innovation and data management practices like data democratization would help organisations achieve competitive advantage. The findings of this study benefit the scholars as it provides future directions for investigating the notion of data-driven innovation and for practitioners to improve their existing data strategy and innovation process.

# 5　References

## Copyright